\documentclass[preprint,authoryear,12pt]{elsarticle}

\usepackage{epsfig}

\usepackage{amssymb}

\usepackage[authoryear]{natbib}

\usepackage[
a4paper=true,%
breaklinks=true,%
colorlinks=true,%
pdfauthor={Zagainova Iu.S., Fainshtein V.G., Obridko V.N.},%
pdftitle={Leading and following sunspots: their magnetic properties and ultra-violet emission above them}%
]{hyperref}

\begin{document}

\begin{frontmatter}



\title{Leading and following sunspots: their magnetic properties and ultra-violet emission above them}


\author[label1]{Zagainova Iu. S.\corref{cor}}
\address[label1]{IZMIRAN (Institute of Earth magnetism, ionosphere and radiowaves propagation named after Nikolay Pushkov of the Russian Academy of Sciences), 142190, Moscow, Troitsk, Russia}
\ead{yuliazagainova@mail.ru}

\author[label2]{Fainshtein V.G.}
\address[label2]{ISTP SB RAS (Institute of Solar-Terrestrial Physics of Siberian Branch 
of the Russian Academy of Sciences), 664033, Irkutsk, P/O Box 291, Russia}
\cortext[cor]{Corresponding author} \ead{vfain@iszf.irk.ru }

\author[label1]{Obridko V.N.}
\ead{obridko@izmiran.ru}


\begin{abstract}
Using SDO/HMI and SDO/AIA data for sunspot groups of the 24th solar cycle, we analyzed magnetic properties and $He~II~304$\AA$~$ emission in leading and following sunspots separately. Simultaneous examination of umbral magnetic properties and atmospheric characteristics above the umbrae draws on average differences in $He~II~304$\AA$~$ contrast over the umbrae of leading and following spots we discovered earlier for solar cycle 23 sunspot groups based on SOHO data as well as on the hypothetical relationship between contrast asymmetry and magnetic field asymmetry in umbrae. We use a more accurate and faster algorithm for solving the pi-uncertainty problem of the transverse magnetic field direction in this research producing new results on differences in magnetic field properties between magneto-conjugated leaders and followers. We found that, in $\approx~78\%$ of the cases, the minimum (over the umbra area) angle between the magnetic field line and the normal to the solar surface, $\alpha_{min}$, is smaller in the leading spots, so the magnetic field there is more vertical than that in the counterpart following spot. It was also found that umbral area-avegared angle $<\alpha>$ in $\approx~83$\% of the spot groups examined is smaller in the leader compared to the follower and the maximum and mean magnetic flux densities inside the umbra depend on the umbral area. Moreover, it was shown that a negative correlation exists between $\alpha_{min}$ and umbral area $S$, as well as between $<\alpha>$ and $S$, maximum magnetic induction $B_{max}$ and $\alpha_{min}$, mean magnetic induction $<B>$ and $<\alpha>$. We also discovered the existence of a positive correlation for the dependence of $B_{max}$ on $S$, and $<B>$ and $S$, differing for leaders and followers. It was shown that a positive correlation exists between $He~II~304$\AA$~$ contrast $C_{304-L}$ and $<\alpha_{L}>$ for leaders and between $C_{304-L}/ C_{304-F}$ and $l_L/l_F$, where $l_L$ ($l_F$) is the length of magnetic field line from leader (follower) to the field line apex. Comparing the contrast to $\alpha_{min}$, maximum magnetic induction $B_{max}$ and magnetic flux $<B>(S)$ in leader and follower umbrae demonstrated an absence of any relationship between contrast and those parameters. In this study, we again confirmed that these results allow us to suggest that the leading sunspot might be formed at deeper levels than the following ones. 

\end{abstract}


\end{frontmatter}

\parindent=0.5 cm

\section{Introduction}

According to photospheric observations, sunspots are characterised by decreased values of material temperature and brightness, as well as by increased values of magnetic field compared to the quiet photosphere \citep{Bray1964,Obridko1985,Maltby1992}. The number of spots simultaneously observed in the Sun is a major characteristic of the solar activity and its cycles \citep{Bray1964,Murdin2000}. The spots are intimately related to the manifestations of other forms of solar activity, e.g. flares.

The emergence and further evolution of sunspots is a rather complex physical process, and the properties of individual spots, on the one hand, can differ significantly, while on the other are closely related to each other and the ambient solar regions both in the subphotospheric layers and at various heights of the solar atmosphere (see, e.g. \citep{Pipin2011} and references therein).

Sunspots often form groups, where spots with differing properties can be found; the group itself having its special characteristics determined by all the group spots combined. In most cases, the westernmost sunspot of the group having a larger area and located closer to the equator compared to the other spots in the group is called the leading or head spot. The sunspots exhibiting the opposite field polarity are called following or tail spots. According to Hale's law of sunspot polarities in a group: ,,...in odd cycles the magnetic field of the leading sunspots in groups in the northern hemisphere has the north polarity, in tail sunspots, the south polarity. This pattern reverses its sign in the southern hemisphere or upon entering the even cycle'' \citep{Obridko1985}.

The overwhelming majority of earlier investigations examined sunspot properties irrespective of their type: leading/following. Papers comparing leading to following sunspot properties in a single group or, averaged, for several groups have been relatively few. It was shown that the area dependences of sunspot emission contrast \citep{Sobotka1986} and photospheric magnetic field in sunspots \citep{Bray1964} exhibit practical no difference between leading and following spots or sunspot evolution stage. \citep{Gilman1985} found a slight difference in rotation speeds between leadings and followings. 

Recent research has shown a noticeable difference in leading/following sunspot properties as observed in different spectral ranges. Thus, it was shown in \citep{Zagainova2011} that the dependence of $He~II~304~$\AA$~$ contrast and $He~I~10830~$\AA$~$ parameters of an infrared (IR) triplet on the umbra area differ considerably for leading/following spots.

Magnetic properties have also been found to differ between leading and following sunspots. Couples of magneto-conjugated leading/following sunspots were identified in \citep{Zagainova2015}, their umbrae connected through magnetic field lines, based on Bd-technique based potential approximation computations of magnetic field \citep{Rudenko2001} as well as SDO data for 2010-2013. In $\sim 81\%$ cases, the minimum angle between the field line and the normal to the solar surface, $\alpha_{min}$, was found to be smaller in the leading than in the following. In other words, the magnetic tube connecting leader and follower umbrae was found, in most cases, to be more radial in the leader than in the follower. Analysis of these case showed that there is a positive correlation between $\alpha_{min}$ for leadings and that for followings. 

The umbral area dependence of angle $\alpha_{min}$ was shown to differ for leadings and followings. A weak negative correlations was found between the $\alpha_{min}$ values and the maximum value of magnetic induction in the umbra. In other words, magnetic field lines are, on average, more radial in magnetic tubes forming the umbrae of both the leading and following spots and having stronger fields at photospheric level. 

It was suggested \citep{Zagainova2011} that differences in the solar atmosphere properties of the leadings' and following's umbrae are caused by the asymmetry of a magnetic tube connecting the leading and following parts. This may result in increasing of the ultraviolet $\lambda 304~$\AA$~$ emission above the following sunspot as compared to the leading spot. This fact in turn can explain the differences in the umbral area dependence of the $He~I~10830$ IR triplet parameters between leadings and followings. This conclusion relies on the idea that the chromospheric helium IR triplet is formed via an ionisation-recombination mechanism. This mechanism involves helium atoms ionised by an ultraviolet range radiation flux, followed by a portion of these atoms entering, after a certain lag, the metastable level, $2^3s$, accompanied by absorbed emission of the photospheric continuum \citep{Nikolskaya1966,Livshits1975,Pozhalova1988}. It was also shown that the properties of single sunspots were the same as those of leadings.

This paper continues the research started in\citep{Zagainova2011,Zagainova2015,Zagainova2015a}, where we used a new method for fast and accurate azimuth disambiguation of vector magnetogram data. We compared magnetic properties of leading and following sunspots found using vector measurements of magnetic field by the high spatial resolution SDO/HMI instrument during the growth phase and maximum of solar cycle 24. For the same time period, observations at $\lambda 304~$\AA$~$ of the Sun by the SDO/AIA instrument \citep{Lemen2012} were used to compare the dependences of $He~II~304$\AA$~$ contrast above umbra on the umbral area of leadings and followings.

\section{Data and research methods}

As in our previous study \citep{Zagainova2015a}, we found umbral magnetic characteristics based on vector magnetograph SDO/HMI data, but the determination of the field vector characteristics involves the procedure of solving an pi-ambiguity problem when finding the direction of the transverse field. In \citep{Zagainova2015a}, we used HMI data, for which the pi-ambiguity problem had been solved. In this paper, we solved the pi-ambiguity problem ourselves using a more accurate and faster method proposed in paper \citep{Rudenko2014}. 

A set of 40 bipolar groups observed during 2010-2012, and a set of 29 single sunspots with regular-shaped penumbra and pores observed in 2010 -2011, were selected for the analysis. Fig. 1 presents examples of magneto-conjugated sunspot couples and a single sunspot. Selected magneto-conjugated sunspots are listed in Table 1, which also show the NOAA active region (AR) number, umbral area of the leading spots $S_L$ and of the following spots $S_F$ in millionths of solar hemisphere (MSH) and the magnetic polarity of the spot (north polarity $N$ or south polarity $S$). Data for single spots are gathered in Table 2. The umbral area was found from sunspot images in continuum based on SDO/HMI data. The AR number was found from data in {http://www.solarmonitor.org/} in solar images obtained by SDO/HMI. Compared to the analagouse Table 1 in \citep{Zagainova2015}, some alterations are made in Table 1 in this paper.

The following criteria were kept in mind when selecting spots for determining the magnetic properties of their umbrae. First, the selected (leader-follower) sunspot pairs must be magneto-conjugated. This means that the field lines coming from the leader umbra must end either in, or near, the follower umbra. Correspondingly, the field lines coming from the follower umbra must end either in, or near, the leader umbra. The presence of field lines coming from the one spot but not ,,hitting'' the other spot is related to a relatively low spatial resolution of potential approximation calculations of the field. Followers were pores in some of the magneto-conjugated spots. The second criterion for the spots to be selected is that they must be close to the central meridian. This requirement is mainly related to poor accuracy when determining the umbral area when sunspots are far from the central meridian (CM), i.e. at an angular distance of more than $30^{\circ}$. In fact, in all cases but one, the sunspots under study never digressed more than $20^{\circ}$ from the CM. And, lastly, the magneto-conjugated sunspots have been found to be characterised by a clear-cut regular-shaped umbra with a circular or elliptic symmetry, while single sunspots must have a clear-cut umbra completely engulfed by the penumbra. 

\small
\begin{table}
\caption{The list of magneto-conjugated sunspot pairs selected for analysis.}
\begin{tabular}{|c|c|c|c|c|c|c|c|c|c|}
\hline 
Date&AR &$S_L$,&$S_F$,&Field &Date&AR &$S_L$,&$S_F$,&Field \\ 
 &NOAA &MSH &MSH &polarity& &NOAA &MSH &MSH &polarity, \\
 &number& & &N/S & &number& & &N/S \\ \hline 
2010.09.27 & 11109 & 51 & 10 & N & 2011.08.07 & 11266 & 3 & 4 & N \\ \hline
2010.10.25 & 11117 & 15 & 22 & N & 2011.08.22 & 11272 & 4 & 2 & S \\ \hline
2010.10.25 & 11117 & 21 & 5 & N & 2011.08.22 & 11272 & 9 & 5 & S \\ \hline
2010.11.29 & 11130 & 9 & 8 & N & 2011.10.15 & 11316 & 20 & 22 & S \\ \hline
2011.01.04 & 11142 & 4 & 5 & S & 2011.10.15 & 11319 & 8 & 3 & N \\ \hline
2011.02.02 & 11150 & 7 & 7 & S & 2011.10.28 & 11330 & 68 & 7 & N \\ \hline
2011.02.13 & 11158 & 12 & 12 & S & 2011.11.01 & 11334 & 15 & 2 & N \\ \hline
2011.02.13 & 11158 & 6 & 9 & S & 2011.11.07 & 11339 & 37 & 29 & N \\ \hline 
2011.03.08 & 11166 & 30 & 21 & N & 2011.11.30 & 11361 & 18 & 10 & N \\ \hline
2011.03.08 & 11166 & 22 & 31 & N & 2011.12.03 & 11365 & 7 & 4 & N \\ \hline
2011.03.11 & 11169 & 23 & 8 & N & 2011.12.03 & 11363 & 27 & 12 & S \\ \hline	
2011.04.01 & 11183 & 19 & 5 & N & 2011.12.05 & 11363 & 9 & 1 & S \\ \hline
2011.04.03 & 11184 & 7 & 3 & N & 2011.12.05 & 11364 & 61 & 15 & N \\ \hline
2011.04.13 & 11190 & 10 & 13 & N & 2011.12.20 & 11382 & 23 & 3 & S \\ \hline
2011.04.18 & 11193 & 21 & 4 & N & 2011.12.25 & 11384 & 71 & 8 & N \\ \hline
2011.04.24 & 11195 & 30 & 23 & S & 2012.01.20 & 11401 & 50 & 16 & N \\ \hline 
2011.06.14 & 11234 & 3.3 & 1 & S & 2012.02.01 & 11413 & 15 & 7 & N \\ \hline
2011.07.30 & 11260 & 22 & 7.3& N & 2012.02.11 & 11416 & 33 & 46 & S \\ \hline
2011.08.01 & 11263 & 54 & 62 & N & 2012.02.11 & 11416 & 13 & 37.6 & S \\ \hline
2011.08.03 & 11263 & 63 & 16 & N & 2012.02.20 & 11422 & 45 & 31 & N \\ \hline
 
\end{tabular}
\end{table}
\normalsize

\begin{figure*}
 \centering
 \includegraphics[width=1\textwidth]{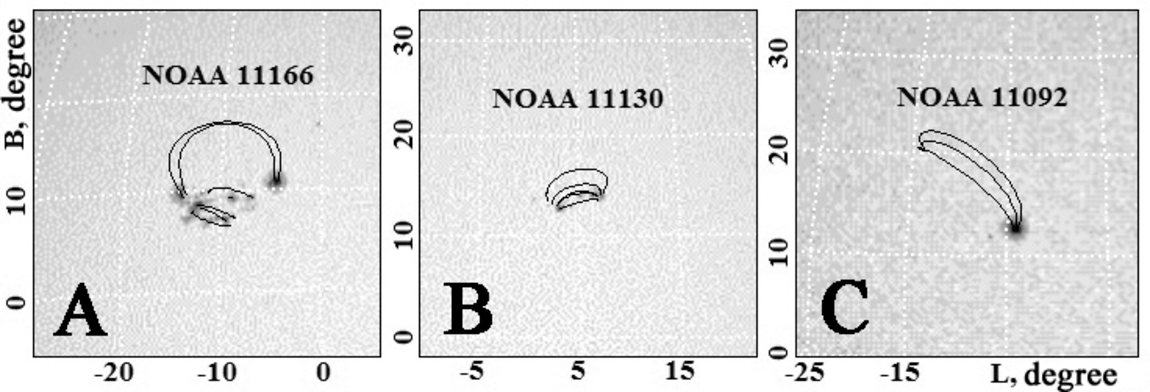}
 \caption{(A) two large magneto-conjugated sunspots with smaller magneto-conjugated or unconjugated sunspots (08.03.2011) in between; (B) a pair of small-scale magneto-conjugated sunspots (29.11.2010); (C) single spot with leader magnetic properties (03.08.2010). The field lines connecting leading and following sunspots are plotted based on potential approximation calculations of magnetic field based on SOLIS magnetograph data, superimposed on SDO/HMI continuum solar images. Field calculations relied on the Bd-technology \citet{Rudenko2001} using field potential decomposition into 90 spherical harmonics. (Adaptation of Fig 1 in \citet{Zagainova2015a}).}
\end{figure*}

\small
\begin{table}
\caption{The list of single sunspots selected for analysis}
\begin{tabular}{|c|c|c|c|c|c|c|c|}
\hline 
Date&AR &$S_S$,&Field &Date&AR &$S_S$,&Field \\ 
 &NOAA &MSH &polarity& &NOAA &MSH &polarity, \\
 &number& &N/S & &number& &N/S \\ \hline 
2010.07.02 & 11084 & 23 & S & 2011.05.22 & 11216 & 16 &	S \\ \hline 
2010.08.03 & 11092 & 42 & N & 2011.06.07 & 11232 & 12 &	N \\ \hline 
2010.08.09 & 11093 & 25 & N & 2011.07.17 & 11251 & 18 &	N \\ \hline 
2010.09.21 & 11108 & 48 & S & 2011.10.08 & 11309 & 18 &	N \\ \hline 
2010.10.19 & 11113 & 22 & N & 2011.10.10 & 11312 & 48 &	N \\ \hline
2010.10.19 & 11115 & 29 & S & 2011.10.10 & 11309 & 14 &	N \\ \hline
2010.10.20 & 11113 & 21 & N & 2011.11.07 & 11338 & 25 &	S \\ \hline
2010.10.20 & 11115 & 25 & S & 2011.11.12 & 11340 & 17 &	S \\ \hline
2010.11.22 & 11127 & 15 & N & 2011.11.12 & 11342 & 30 &	N \\ \hline
2010.12.08 & 11131 & 73 & N & 2011.11.12 & 11341 & 18 &	N \\ \hline
2010.12.09 & 11131 & 64 & N & 2011.11.12 & 11343 & 28 &	N \\ \hline
2011.01.05 & 11140 & 28 & N & 2011.11.17 & 11346 & 15 &	S \\ \hline
2011.03.30 & 11180 & 6 & N & 2011.11.24 & 11355 & 23 & N \\ \hline
2011.04.10 & 11185 & 4 & N & 2011.11.26 & 11360 & 19 & N \\ \hline
2011.05.05 & 11203 & 21 & N & - & - & - & - \\ \hline
					
\end{tabular}
\end{table}
\normalsize

This paper analysed magnetic field characteristics in sunspot umbra, such as: the minimum angle $\alpha$ between the field direction and the normal to the solar surface at the field measurement point ($\alpha_{min}$) (see Fig.2 for details on how angle $\alpha$ was found; angle $\alpha_{min}$ is the smallest of the $\alpha$ angles as calculated at different points where umbral magnetic field was measured); umbral area-averaged (denoted $<~>$) angle $\alpha$: $<\alpha>$; maximum $B_{max}$ and the average value of magnetic induction $<B>$ within the umbra. Note, that when the magnetic field had a negative polarity - the field vector directed sunwards - angle $\alpha$ was found to exceed $90^{\circ}$, therefore, when comparing this angle to the angles for a positive field polarity (field vector sunward), the values of $\alpha$ were subtracted from $180^{\circ}$: $\alpha_{min}= min(180^{\circ}-\alpha)$, $<\alpha>=<(180^{\circ}-\alpha)>$.

Magnetic field properties in umbrae were analysed using vector magnetic field measurements by the HMI magnetograph (\citet{Scherrer2012}; {http://hmi.stanford.edu/}) which allow one to determine magnetic induction $B$; magnetic field vector tilt to line-of-sight $\delta$; and azimuth $\Psi$, measured in the plane of the sky from CCD array column, counter-clockwise, to the magnetic (transverse) field vector as projected on this plane. In magnetic field measurements, the spatial resolution of the HMI instrument is $\approx 0.5''$. Our analysis involved magnetograms that were closest to the time moments when the continuum solar images were obtained.

To be able to find angle $\alpha$ based on measured values, we obtained ratios between angles $\delta$, and $\Psi$. The calculations were in the Descartes system of coordinates ($X,~Y,~Z$) centered at solar disk centre, where the $OX$, $OY$ axes are in the plane of the sky, and the $OY$ axis passes through the North pole (the angle between the elliptical plane and the equatorial plane is ignored). The $OZ$ axis is directed along the line of sight and is perpendicular to the plane of the sky. The line of sight is assumed to be perpendicular to the plane of the sky at all points within the solar disk. 

The angle between the magnetic field direction and the radial direction was found from: $\cos(\alpha)=B_r/ B$. 

\begin{eqnarray}
\mathbf B_r={\mathbf B_X} X+{\mathbf B_Y} Y+{\mathbf B_Z} Z &=&
B\sin{\delta}\cos(\Psi+90^\circ)+{} \nonumber\\
B\sin{\delta}\sin(\Psi+90^\circ)+B\cos{\delta}
\end{eqnarray}

Angle $\delta$ is set between $\mathbf B$ and the $OZ$ axis as found from SDO data, $\delta=[0^{\circ}; 180^{\circ}]$ as well as angle $\Psi$, between the $OY$ axis and the projection of $\mathbf B$ onto the $XY$ plane (azimuth in SDO data), $\Psi=[0^{\circ}; 360^{\circ}]$, counter-clockwise from the $OY$ axis. Our calculations ignored the angle between the ecliptic and the solar equator.

\begin{figure*}
 \centering
 \includegraphics[width=0.7\textwidth]{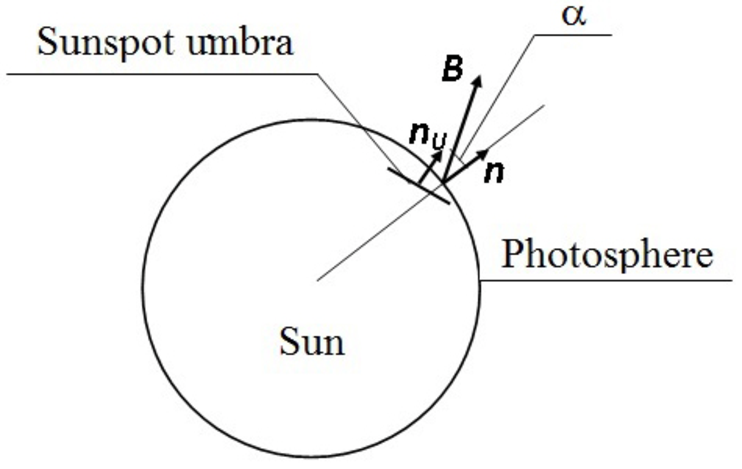}
 \caption{An illustration of the angle under analysis, $\alpha$, between the magnetic field vector $\mathbf B$ and the positive normal $\mathbf n$ to solar surface at the field measurement point. $\mathbf n_{u}$ is the positive normal to the spot umbra. In the general case, the umbra is assumed to be non-perpendicular to the radial direction from the solar centre.}
\end{figure*}

All plots using data on angles $\alpha_{min-L}$, $\alpha_{min-F}$, and/or maximum magnetic induction $B_{max-L}$, $B_{max-F}$, were drawn for leaders and followers satisfying the condition: $\alpha_{min-L} \le \alpha_{min-F}$. Correspondingly, plots with data onf umbral area-averaged angles $<\alpha_L>$, $<\alpha_F>$, and/or average magnetic induction $<B_L>$, $<B_F>$, were drawn for spots with $\alpha_{min-L} \le \alpha_{min-F}$.

Our analysis of solar emission in the $\lambda 304~$\AA$~$ line above sunspot umbrae relied on SDO/AIA telescopic data \citep{Lemen2012}. This telescope provides spatial resolution $\approx 0.5''$. $\lambda 304~$\AA$~$ line contrast above umbrae $C_{304}$ is found from $C_{304}=I_s/I_0$, where $I_s$ is intensity measure in the umbra, $I_0$ - in the quiet area (for details, see \citep{Zagainova2011}). The umbral area as expressed in MSH was found from spot images in continuum based on SDO/HMI data. 

\section{Results}
\subsection{Comparison between magnetic properties of leading and following sunspots during the growth phase and maximum of solar activity cycle 24} 

Vector measurements of magnetic field by SDO/HMI were used to identify and compare magnetic field properties in the umbrae of leading and following spots. These measurements had a considerably higher spatial resolution than what was provided by potential approximation calculations of magnetic field in the solar atmosphere that served as a basis for a similar analysis of field properties in leadings and followings in our previous paper \citep{Zagainova2015}.

We will start to compare magnetic properties in leader and follower umbrae by comparing the minimum ($\alpha_{min}$) and average ($<\alpha>$) angles between the magnetic induction direction and the radial direction at field measurement point in the umbrae of the two types of spots. It was shown earlier in our papers based on various data that in $\approx 80\%$ \citep{Zagainova2015} and $\approx 84\%$ \citep{Zagainova2015a} of the magneto-conjugated sunspot pairs $\alpha_{min}$ was found to be larger in followers than in leaders. In other words, the magnetic tube connecting leader and follower umbrae was found, in most cases, to be more radial in the leader than in the follower. 

Angles $\alpha_{min}$ and $<\alpha>$ as compared here between leader and 
follower umbrae has shown that in $\approx 78\%$, $\alpha_{min-L}\le \alpha_{min-F}$, whereas in $\approx 83\%$, $<\alpha_L>\le <\alpha_F>$. For the spot groups in question (leader/follower) meeting these requirements, sample-averaged angle $\alpha_{min-L-av}$ in leaders $\approx 2.14^{\circ}$, in followers $\alpha_{min-F-av}\approx 10.17^{\circ}$, with $\alpha_{min-L-av}/ \alpha_{min-F-av}\approx 0.21$. For average values of $<\alpha>$ we have $<\alpha_{L-av}>$ in leaders $\approx 30.45^{\circ}$, in followers $<\alpha_{F-av}>\approx 38.29^{\circ}$, and $\alpha_{min-L-av}/ \alpha_{min-F-av} \approx 0.8$.

It was shown in \citep{Zagainova2015} and \citep{Zagainova2015a} that a positive correlation exists between the $\alpha_{min-F}$ and $\alpha_{min-L}$ values for leaders and followers with $\alpha_{min-L}\le \alpha_{min-F}$. Our new analysis has demonstrated that correlation between these angles persists but the correlation coefficient has fallen from $r=0.77$ to $r=0.48$, Fig. 3(A). {\bf It has been found that spots satisfying the condition $<\alpha_L>\le <\alpha_F>$ also exhibit a positive correlation between $<\alpha_L>$ and $<\alpha_F>$, with coefficient $r\approx 0.45$, Fig. 3(B).}

\begin{figure*}
 \centering
 \includegraphics[width=0.95\textwidth]{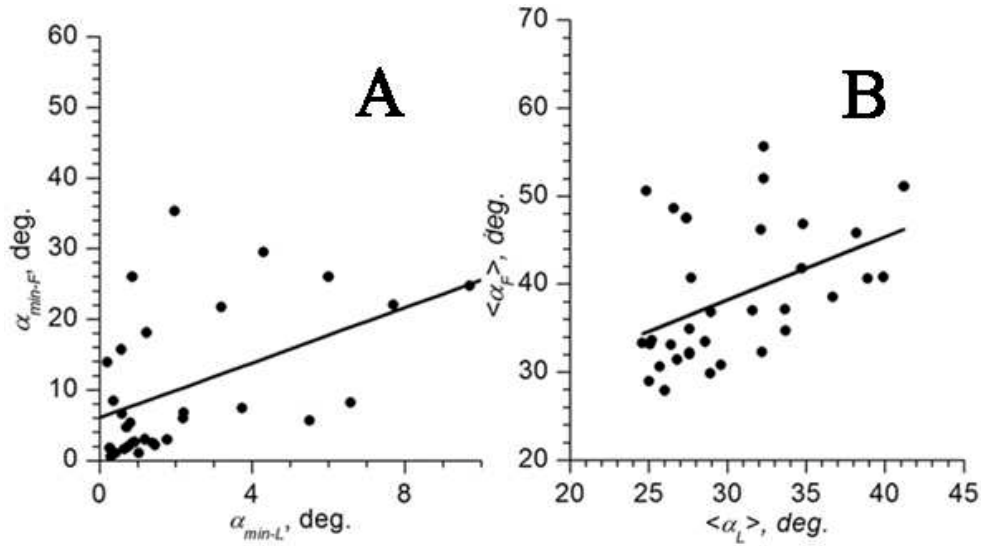}
 \caption{Relationship between minimum angles between magnetic field direction and radial direction in the umbrae of leadings, $\alpha_{min-L}$, and followings, $\alpha_{min-F}$ (A; $\alpha_{min-F}(\alpha_{min-L})=6.03+1.95\alpha_{min-L}$, correlation coefficient $r=0.48$) and mean values of angles $\alpha$ in leaders ($<\alpha_L>$) and followers ($<\alpha_F>$) (B; $<\alpha_F>(<\alpha_L>)=16.64+0.72<\alpha_L>$, $r=0.45$).}
\end{figure*}

In relation to these findings, the question arises why $\alpha_{min}$ is not $0$, but a few degrees or more instead? A possible explanation is given in \citep{Kuklin1985}, where it is argued that an observable spot umbra may not be perpendicular to the radial direction from the solar centre. In this case, the minimum angle between the field direction an the normal to the sunspot umbral plane can be close to $0$, but the minimum angle between the field direction and the normal to the solar surface will differ from $0$, being a few degrees or more. This means that the angle $\alpha_{min}$ can also be regarded as the measure of the inclination between the normal to the observable umbral surface and the radial direction. 

Our conclusion that, for most magneto-conjugated leader-follower sunspot pair we studied the magnetic field is more vertical in leading spots than in followers in most cases, does not agree with findings in several theoretical investigations and a number of observations. The ascent of a magnetic tube from the depths of the convection zone to the solar surface was theoretically investigated in \citep{Fan1993,Caligari1995}. It was suggested that the legs of exactly such tubes form sunspots. It was shown in both the papers that there is asymmetry, in the inclination to solar surface, between the western and eastern legs of the tube, that, in the photosphere, manifest themselves as a leader and follower spot. According to the above calculations, the eastern leg is more vertical (i.e. closer to the radial direction) than the western leg. \citet{Van Driel-Gesztelyi1990} presented observation results presumably supporting the conclusion about the field asymmetry between leaders and followers. 

It remains unclear how the resulting contradiction could be solved. What can be said with certainty is that it is not related to the accuracy with which $\alpha_{min}$ was determined, because we used two methods to find this angle: potential approximation calculations of the magnetic field \citep{Zagainova2015} and based on SDO/HMI data (\citet{Zagainova2015a} and this paper). Spatial resolution of calculations of the magnetic field distribution at photospheric level was found, in the former case, at a spatial resolution that was one order lower than the HMI instrument. As a result, on average the $\alpha_{min}$ values for both leaders and follower spots were áîëüøå in field calculations compared to the values found from HMI. However, the conclusion that field lines in the leading sunspots are more radial than in the following in most of the magneto-conjugated spot pairs under study proved to be true in both cases! Based on our findings, we think that the conclusion that a large number of magneto-conjugated leaders and followers exist in which the magnetic tube section from the leader is more radial than in the follower is reliable and supported by observations and calculations of magnetic field configuration at photospheric level. 

As we see it, the likeliest reason for the difference in theoretical analysis results regarding the inclination of the western and eastern legs of the emerging magnetic loop and field line inclination angles in the leader and follower umbrae is a different mechanism for leader and follower formation than the ascent of a magnetic tube through the convection zone. This is indirectly evidenced by the examples of field lines from the leader and follower umbrae and from a single spot (Fig. 4). Part of the field lines can be seen to connect leader and follower, while some field lines from both spots go in various directions spanning different distances. Field lines from a big single sunspot are distributed rather chaotically. Note that the examples in Fig. 4 are not typical. In most cases, the spread of field line directions from spot umbrae is much smaller.

That such a spread does exist is demonstrated by the following analysis. On average, the umbral area and the mean and maximum magnetic induction in the umbrae of magneto-conjugated spots is higher in the leader than in the follower. This means that magnetic flux from the leader is also larger than the flux entering the follower. Consequently, in most cases, part of the flux from the leader umbra must close, not in its respective follower, but either in other followers, or in another active region, or in other parts of the photosphere. This was observed for many of the sunspots under investigation.
 
A similar analysis can also be done for large single spots, their outgoing magnetic fluxes capable of travelling in various directions. Therefore, to clarify the true ratio between the field line inclination angles to the radial direction for leader and follower spots requires inspecting the emergence from the convection zone of not one magnetic tube, but of a complex magnetic structure consisting of several tubes coming from the umbra of a large leader spot. 

\begin{figure*}
 \centering
 \includegraphics[width=0.95\textwidth]{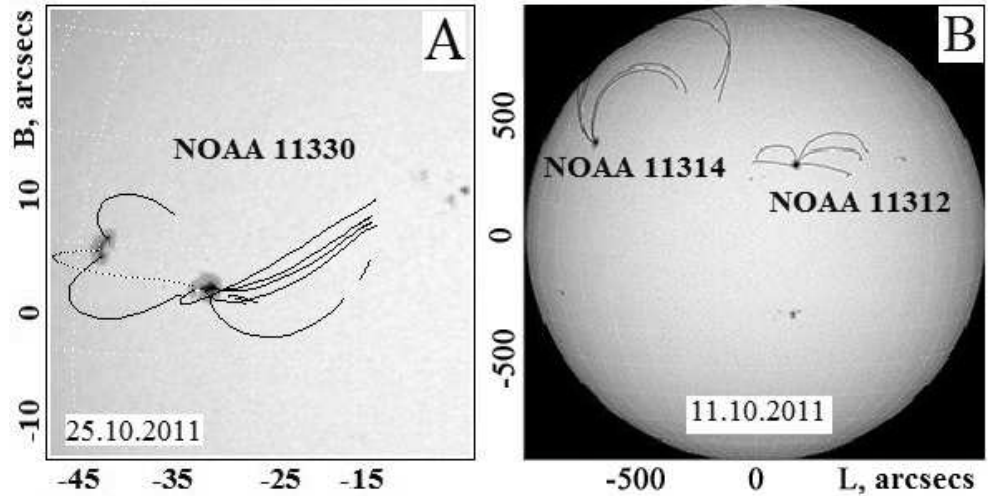}
 \caption{(A) an example of magneto-conjugated leaders and followers when part of the magnetic flux from the leader umbra goes, not into the follower, but into other parts of the photosphere (AR NOAA11330). {\bf (B) an example of a single spot sprouting field lines in various directions (AR NOAA 11312).}}
\end{figure*} 

Another reason for the above difference as to how field line inclination in leaders and followers are related, based on theoretical calculations, relies on the ,,not deep'' sunspot concept \citep{Solovev2014}. In this case, the spots are formed not by the loop ,,legs'' going deep into the convection zone, but by short magnetic ,,columns'' where one can ignore the effect of the Coriolis force on both them and the magnetic tube connecting them. 

Finally, the relation between the field line inclination angles in the leader and follower may depend on sunspot evolution stage. To find out whether this relation exists or not would require dedicated and rather laborious investigations, which are outside the scope of this paper. {\bf We will highlight only some difficulties to be solved for this problem. First, now there are no generally accepted criteria for separating evolution phases of the active region containing magneto-conjugated leaders and followers. This is associated in particular with the diverse nature of time variations in different characteristics of the sunspot umbra: area, maximum and mean magnetic induction within the umbra, etc. (see, e.g., \citep{Cowling1946}). Duration of phases of emergence, smooth evolution and disappearance of leaders and followers is different. However, for a small sampling (9 events) of the magneto-conjugated sunspot pairs considered, we have carried out such a study. Our preliminary analysis showed that the correlation between inclination angles in leaders and followers is most likely to be independent of the evolution phase of the active region.}

It was concluded in \citep{Zagainova2015} and \citep{Zagainova2015a} that both leaders and followers exhibit practically no correlation between minimum angle $\alpha_{min}$ and umbral area $S$ (absolute values for the correlation coefficients for these dependences never exceeded 0.12). In this paper, it was possible to obtain dependences $\alpha_{min}(S)$ and $<\alpha>(S)$, demonstrating a noticeable negative correlation exists for both types of spots, Fig. 5. There is also a relationship between the ratios $\alpha_{min-L}/\alpha_{min-F}$ and umbral areas in leaders and followers $S_L/S_F$, Fig. 5(E), as well as between $<\alpha_L>/<\alpha_F>$ and $S_L/S_F$, Fig. 5(F). Plots in Fig. 5 demonstrate that there is a correlation between the individual parameters for each type of spots (leader/follower) and between two characteristics of asymmetry in leaders and followers.

\begin{figure*}
 \centering
 \includegraphics[width=0.82\textwidth]{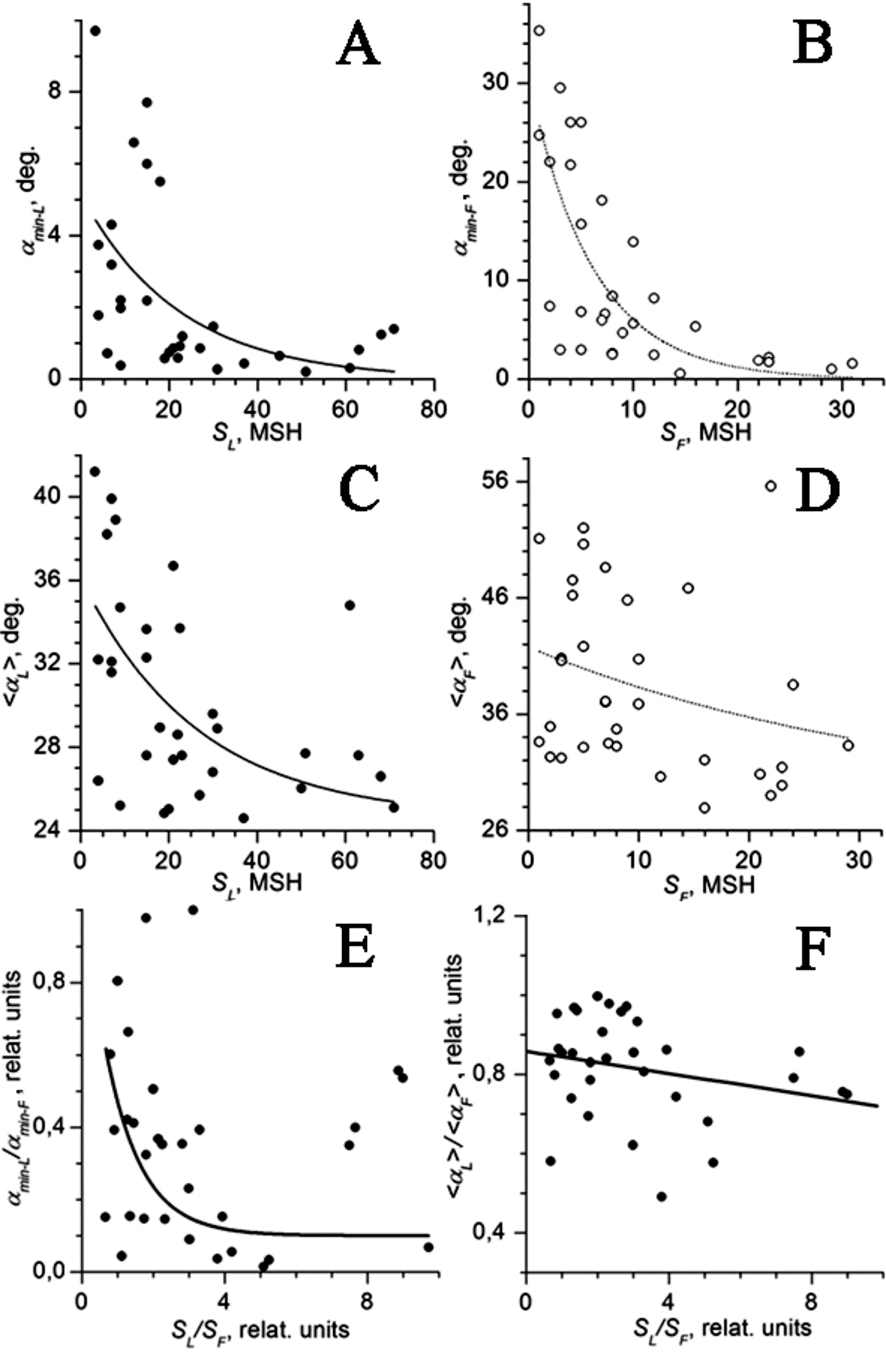}
 \caption{{\bf The minimum angle $\alpha_{min-L}$ (A) and the average angle $<\alpha_L>$ (C) in the leading sunspot umbra depending on the area of its umbra $S_L$. (B, D) - analogous dependencies for followers. (E, F) is the relation $\alpha_{min-L}/\alpha_{min-F}$ and $<\alpha_L>/<\alpha_F>$ vs $S_L/S_F$. The equation of regression line and the correlation coefficient for the dependence (A): $\alpha_{min-L}(S_L)=5.16exp(-S_L/22.2)$, $r=-0.51$: for (B): $\alpha_{min-F}(S_F)= 30.15exp(-S_F/6.22)$, $r=-0.73$; for Ñ: $<\alpha_L>(S_L)=11.54exp(-S_L/26.5)+24.6$, $r=-0.52$;  for (D): $<\alpha_F>(S_F)=13.87exp(-S_F/34.88)+27.9$, $r=-0.29$); for (E): $\alpha_{min-L}/\alpha_{min-F}(S_L/S_F)=0.59exp(-(S_L/S_F)/1.31)+0.13$, $r=-0.25$; for (F): $<\alpha_L>/<\alpha_F>(S_L/S_F)=0.86-0.014(S_L/S_F)$, $r=-0.26$.}}
\end{figure*} 

It was also concluded in \citep{Zagainova2015} and \citep{Zagainova2015a} that magnetic field characteristics in leader/follower umbrae $B_{max-L,F}$ are practically unrelated to angles $\alpha_{min_L,F}$. Our new analysis has demonstrated that leader spots exhibit a negative correlation both between $B_{max-L}$ and $\alpha_{min_L}$, and between $<B_L>$ and $<\alpha_L>$, Fig. 6(A, B). {\bf For followers, the correlation coefficients between these field parameters for spots with $\alpha_{min-L}\le \alpha_{min-F}$ and $<\alpha_L>\le <\alpha_F>$ are, respectively, $r=-0.2$ and $r=-0.12$, i.e. there is practically no relationship between these field parameters.}

\begin{figure*}
 \centering
 \includegraphics[width=0.95\textwidth]{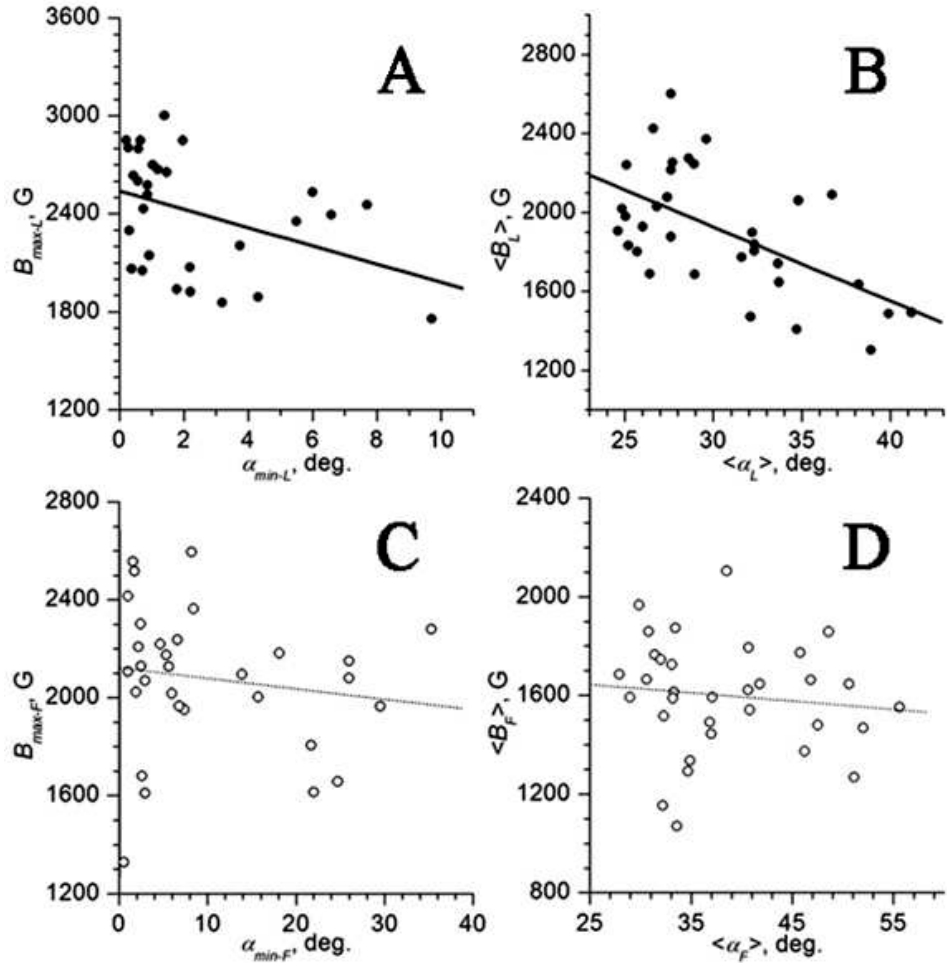}
 \caption{The dependence of $B_{max}$ on $\alpha_{min}$ and of mean magnetic induction $<B>$ on mean angle $<\alpha>$ for leaders (A, $B_{max-L}(\alpha_{min-L})=2539.15-55.82\alpha_{min-L}$, $r=-0.4$; B, $<B_L>(<\alpha_L>)=3051.75-37.49<\alpha_L>$, $r=-0.57$) and followers (C, $B_{max-F}(\alpha_{min-F})=2121.87-4.28\alpha_{min-F}$, $r=-0.2$; D, $<B_F>(<\alpha_F>)=1728.03-3.35<\alpha_F>$, $r=-0.12$).}
\end{figure*}

We discovered a correlation between average angle $<\alpha_{L,F}>$ and $<B_{L,F}>S_{L,S}$ in leaders and followers, Fig. 7. $<B_{L,F}>S_{L,S}$ is regarded as a measure of magnetic flux in umbra $F_{L,F}$.

\begin{figure*}
 \centering
 \includegraphics[width=0.95\textwidth]{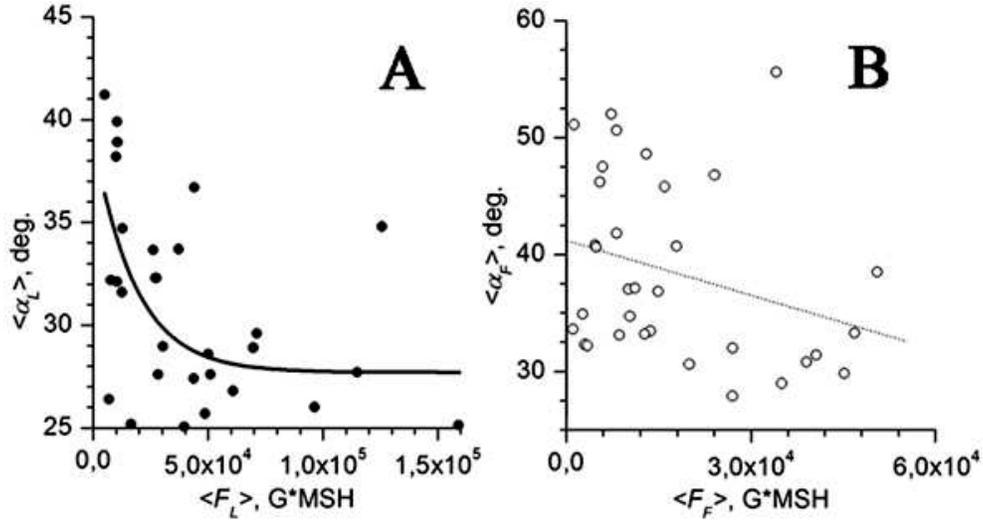}
 \caption{Average angle between field direction and radial direction from solar centre $<\alpha_{L,F}>$ in umbra depending on magnetic flux in umbra $F_{L,F}=<B_{L,F}>S_{L,S}$, for leaders (A, $<\alpha_L>(<F_L>)=27.7+11.42exp(-<F_L>/18221.03)$, $r=-0.59$) and followers (B, $<\alpha_F>(<F_F>)=41.19-1.56<F_F>$, $r=-0.3$).}
\end{figure*} 

The dependence of magnetic induction in spot umbra on umbral area was discussed in a number of papers (see \citet{Ringnes1960} and an overview in the monograph by \citet{Bray1964}). Based on findings of several researches, the conclusion was made that the relationship between maximum magnetic induction $B$ and umbral area $S$ is consistent with the empirical relation obtained in \citep{Houtgast1948} $B=3700S/(S+66)$; here $B$ is measured in G, $S$ in MSH. The relationship between $B$ and $S$ was also studied based on magnetic field vector measurement data in sunspots \citep{Jin2006}. It was shown that there is a logarithmic dependence between the maximum field in a spot in upper layers and its area. At the same time, all the investigations of the $S$ dependence of $B$ never distinguish between leadings and followings.

\begin{figure*}
 \centering
 \includegraphics[width=0.95\textwidth]{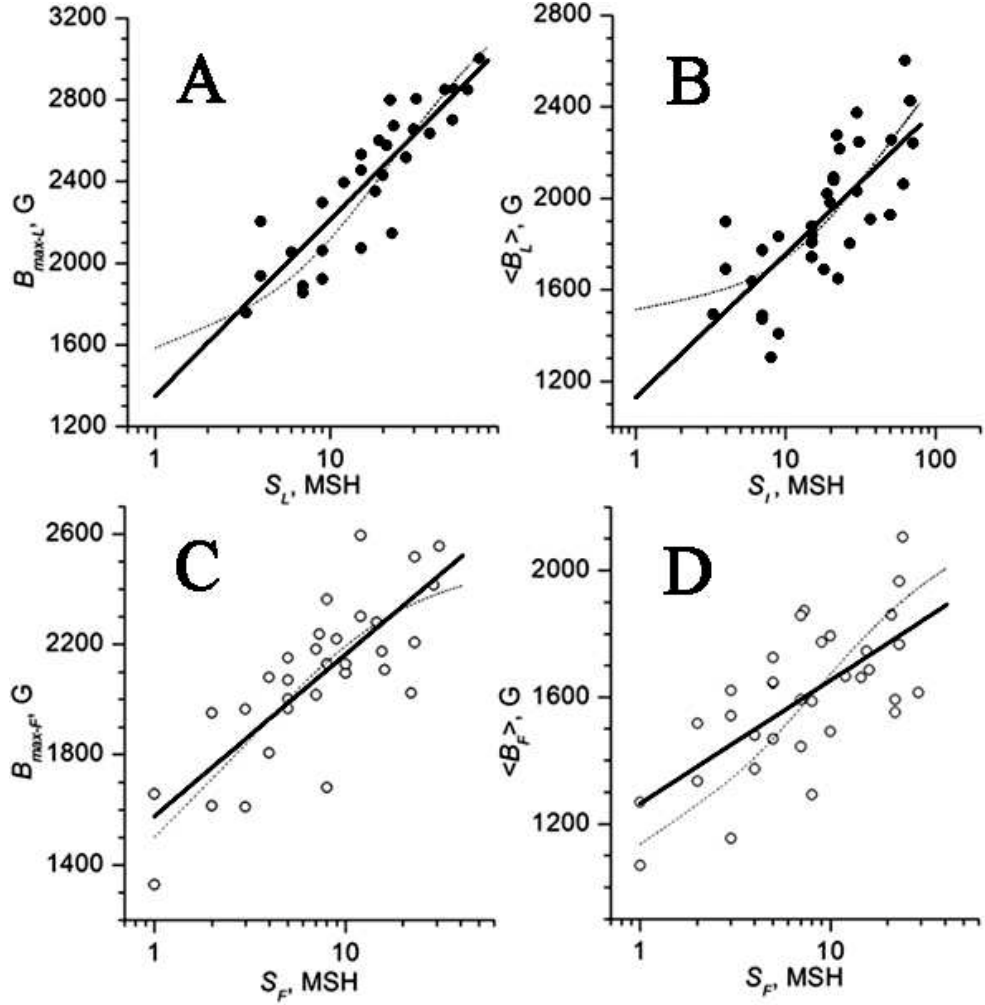}
 \caption{{\bf The dependence of magnetic field value on the area of sunspot umbrae. (A) Maximum field value for leadings. Dotted line: $B_{max-L}(S_L)=1500+2000S_L/(S_L+22.4)$ ($r=0.89$). Solid line: $B_{max-L}(S_L)=375.35\ln{(S_L)}+1350.28$ ($r=0.88$). (B) Mean magnetic field value for leadings. Dotted line: $<B_L>(S_L)=1483+1528S_L/(S_L+50)$ ($r=0.76$). Solid line: $<B_L>(S_L)=272.29\ln{(S_L)}+1129.9$ ($r=0.75$). (C) Maximum magnetic field value for followings. Dotted line line: $B_{max-F}(S_F)=1168+1337S_F/(S_F+3.02)$ ($r=0.8$). Solid line: $B_{max-F}(S_F)=254.56\ln{(S_F)}+1577.17$ ($r=0.79$). (D) Mean field value for followings. Dotted line: $<B_F>(S_F)=1000+1200{S_F}/(S_F+7.82)$ ($r=0.67$). Solid line: $<B_F>(S_F)=169.02\ln{(S_F)}+1264.02$ ($r=0.68$).}}
\end{figure*}

The formula by \citep{Houtgast1948}, contains a very important fault. According to this formula the sunspot magnetic field becomes zero as the spot area tends to zero. It reflects the early period of sunspot research, when it was assumed that there was no magnetic field outside sunspots, the magnetic field varying from $4000~$G in larger sunspots to $100~$G in the smallest spots \citep{Ringnes1960}. With increased quality of observations, however, it became clear that magnetic fields are large enough even in smaller spots. \citep{Steshenko1967,Bumba1967,Beckers1968} demonstrated that the field is no less than $1200~$G, sometimes exceeding $1800~$G in the smallest pores\citep{Baranov1974}. This disadvantage of Houtgast and van Sluiters' formula was also criticised by \citep{Antalova1991,Solovev2014}. 

About the same time, in the 1960s, a concept of ,,kilogauss tubes'' appeared \citep{Sheeley1966,Sheeley1967,Harvey1969,Harvey1971,Livingston1969,Stenflo1973}, according to which, very small formations are possible, with up to $2000~$G strength. At the same time, Houtgast and van Sluiters' formula reflects a very important property – starting from a certain moment, saturation sets in and the dependence of magnetic field on area decreases dramatically.

At the same time, all the investigations of the $S$ dependence of $B$ never distinguish between leaders and followers. For the first time, we compared the maximum and mean values of magnetic induction $B_{max-L,F}$ and $<B>_{L,F}$ separately, in the umbrae of leaders and followers to the area $S_{L,F}$ of the umbrae of these sunspots, Fig. 8. Each plot in this Figure contains two regression lines: one described by the $B_{max}=A+B\ln{(S)}$ function, the other by $B_{max}=A+B{S}/(S+C)$. Using a regression line described by a logarithmic function is evidence that such a function is a good fit for the above point scatters consistent with a similar conclusion in earlier papers (see e.g., \citet{Jin2006}). The regression line described by $Y=A+BX/(X+C)$ ,,resembles'' Houtgast and van Sluiters' formula \citep{Houtgast1948}, but for an important distinction: when $S \longrightarrow {0}$, $B_{max}$ does not tend to $0$. It is to demonstrate this distinction, characteristic for real dependences $B_{max}(S)$, that we include this regression line, which is described by such a formula.

One can see from the plots in Fig. 8 and the approximation formulae that:

\begin{enumerate}
\item Neither the maximum nor average magnetic field drop as low as zero when the area decreases to very small values. 
\item In all cases, magnetic field in leading and single sunspots is larger than in followings.
\item Of much importance is the threshold value of the area triggering curve saturation. In Houtgast and van Sluiters' formula, this value was $66~$MSH and referred to a sunspot of $\approx 8000~$km in radius. Our approximation produces threshold values of the area of $\sim 25~$MSH for leadings and $\approx 10~$MSH for followings. This corresponds to radii of $5000$ and $3100~$km. It could be assumed, conditionally, that saturation sets in when the area is such that the radius is comparable to spot depth. Our findings are on average consistent with estimates by Solov'ev and Kirichek \citep{Solovev2014} thus supporting their concept of a ,,not deep'' sunspot. With such an interpretation, our data indicate that not only do followings exhibit a smaller magnetic field, but may possibly be less deep formations.
\item For very big areas, the asymptotic values are $3550~$G for leadings and $2750 \quad~$G for followings, the smallest possible values being $\approx 1000~$G. 
\item We also compared the mean values of magnetic induction $<B>$ in leading and following umbrae depending on the umbral area. $<B>(S)$ has been found to exhibit more expressed differences between leadings and followings than $B_{max}(S)$. This indicates a more dramatic decrease in magnetic field from the umbral nucleus towards the penumbral boundaries in followings than in leadings.
\item Both for leaders and followers, the logarithmic fit proved to be a good approximation for the $B_{max-L,F}(S_{L,F})$ and $<B_{max-L,F}>(S_{L,F})$ dependences.
\end{enumerate}

We also analysed the magnetic properties of regular-shaped single sunspots with expressed umbra and penumbra and single pores, Table 2. For these, the average value of the minimum angle between the magnetic field direction and the positive normal to solar surface, $\alpha_{min-S-av}\approx 2.57^{\circ}$. It was found that average value $\alpha_{min-S-av}$ is less than average value of $\alpha_{min-L}$ $=\alpha_{min-L-av}=3.32^{\circ}$. At the same time for sunspots with north polarity, the angle has been found to be larger, $\alpha_{min-S-av}-N=2.25^{\circ}$, than for spots with south polarity, $\alpha_{min-S-av}-S=1.73^{\circ}$. The average angle for the single spots under study, $<\alpha>_{av}$, is $29.25^{\circ}$. It was found for single sunspots that $B_{max-S-av} \approx 2834~$G, and the average area of such sunspots $S_{S-av}\approx 25.5~$MSH. This means that compared to leadings ($B_{max-S-av}=2394~$G; $S_{S-av}=24.3~$MSH), the single sunspots selected for analysis are characterised by larger values of maximum magnetic field and larger area of the umbra. This supports the conclusion in \citep{Zagainova2015a} and in this paper that sunspots with stronger magnetic field in the umbra and larger area are, on average, characterised by smaller minimum angles $\alpha_{min}$. There is practically no relationship between the minimum angle in single sunspots, $\alpha_{min-S}$, and maximum magnetic induction $B_{max-S}$. A weak negative correlation exists between $<\alpha_S>$ and $<B_S>$ ($r=-0.26$). {\bf A negative correlation has been found to exist between $<\alpha_S>$ and $S_S$, with correlation coefficient $r\approx -0.66$, Fig. 9(C).}

\begin{figure*}
 \centering
 \includegraphics[width=1.0\textwidth]{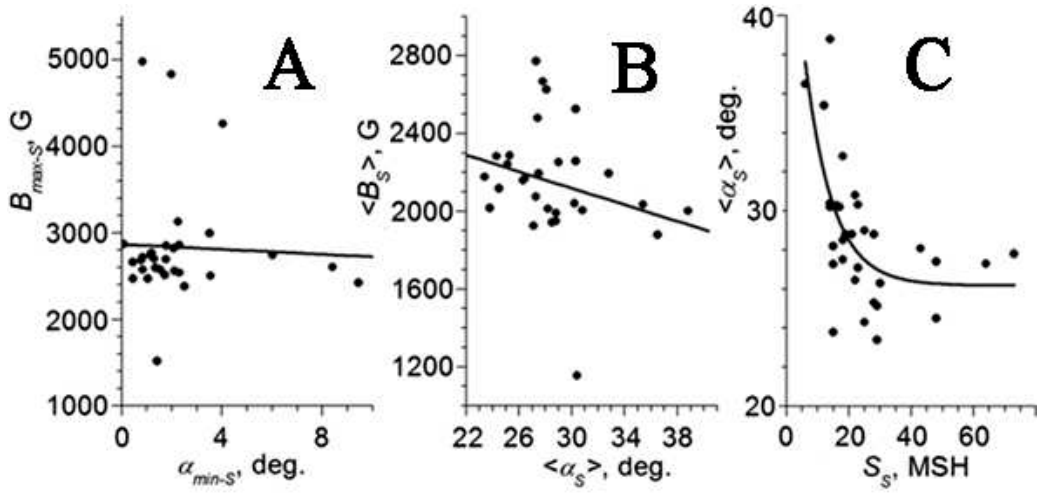}
 \caption{{\bf The dependencies for single sunspots: (A) of the magnetic induction maximum in the sunspot umbra $B_{max-S}$ on the minimum angle in the umbra $\alpha_{min-S}$; (B) of the average magnetic induction in the sunspot umbra $<B_S>$ on the average one in the umbra of the angle $<\alpha_S>$; (C) of $<\alpha_S>$ on the sunspot umbra area $S_S$. The equation of regression line and the correlation coefficient for the dependence (A): $B_{max-S}(\alpha_{min-S})=2866.58-14.11\alpha_{min-S}$, $r=-0.04$; for (B): $<B_S>(<\alpha_S>)= 2754.63-21.2<\alpha_S>$, $r=-0.26$; for (C): $<\alpha_S>(S_S)=22.7exp(-S_S/8.75)+26.19$, $r=-0.66$.}}
\end{figure*} 
 
\subsection{Comparing helium $\lambda 304~$\AA$~$ emission contrast above leader and follower umbrae to their magnetic properties}

It was shown in \citep{Zagainova2011}, for solar cycle 23, that $He~II~304~$\AA$~$ contrast is different above leader and follower umbrae: on average, contrast in this line is higher in followers compared to leaders or single spots. This result was later confirmed \citet{Zagainova2015a} for solar cycle 24 growth stages and maximum (see Fig. 9). It has been found that, for both the periods under study, $He~II~304~$\AA$~$ contrast above leaders and followers, on average, practically does not depend on umbral area. 

\begin{figure*}
 \centering
 \includegraphics[width=0.95\textwidth]{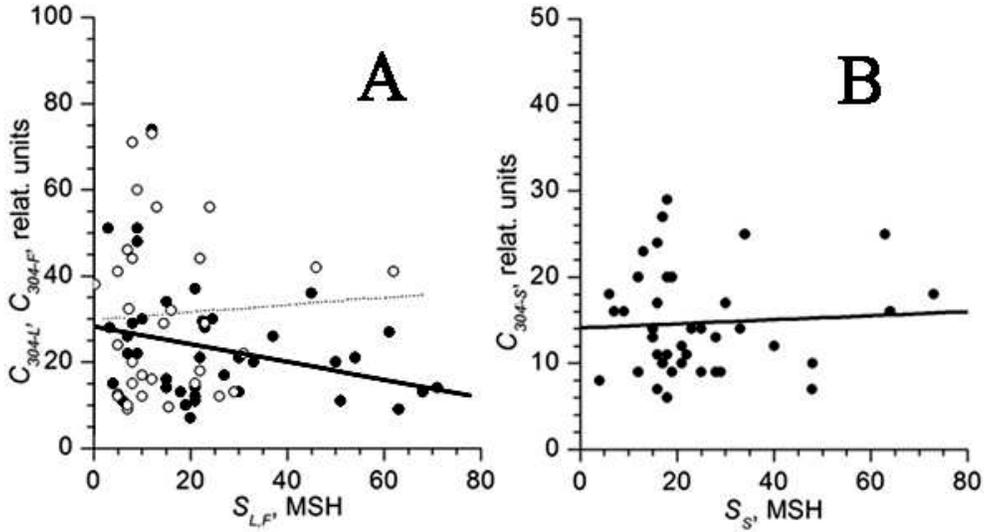}
 \caption{$He~II~304~$\AA$~$ contrast above sunspot umbrae vs their area. (A) for leaders and single spots (disks $r=-0.3$ and solid $r=-0.06$ regression line) and followers (circles and dotted regression line; $r=0.06$), (B) for single spots ($r=0.06$). (It is an adaptation of Fig 2 in \citet{Zagainova2015a}).}
\end{figure*}

The dependences in Fig. 10 can be interpreted as follows: the magnetic field line configuration in sunspot groups is such that more $He~II$ ions emitting in $\lambda 304~$\AA$~$ are accumulated above the following umbrae than above the leading umbrae. \citet{Zagainova2011} suggested that different UV emission fluxes may be due to the asymmetry of magnetic tubes connecting leaders and followers. This asymmetry, in turn, may result in magnetic field properties differing between the umbrae of the two types of spots. It was shown in this paper, as well as in our earlier paper \citep{Zagainova2015,Zagainova2015a}, that this difference in the magnetic properties of the umbrae of two types of sunspots, as well as the asymmetry in magnetic field lines connecting leaders and followers do exist. The maximum and mean values of magnetic induction differ between leader and follower umbrae, while the field itself is found, in most cases, to be more vertical in leader umbrae than in follower umbrae. In many cases magnetic field lines connecting the two types of spots are also found to be nonsymmetrical: the smaller the minimum angle between the field line in the sunspot and the normal to the solar surface, the shorter the field line section between the spot and the apex of the line.

At the same time, findings in \citep{Zagainova2015,Zagainova2015a} failed to provide complete answers to the following questions: is there a quantitative relation between $He~II~304~$\AA$~$ contrast above leader ($C_{304-L}$) and follower ($C_{304-F}$) umbrae and umbral magnetic properties of these two types of spots? Is there a quantitative relation between the characteristics of $He~II~304~$\AA$~$ contrast asymmetry $C_{304-L}/C_{304-F}$ and the characteristics of the asymmetry in leader and follower magnetic properties (for example, $B_{max-L}/B_{max-F}$; $\alpha_{min-L}/\alpha_{min-F}$ and others)? In this section, we will discuss in more detail the dependence of $C_{304-L}$ and $C_{304-F}$ on magnetic field characteristics in leader and follower umbrae. 

Our analysis revealed practically no correlation between $C_{304-L}$ and $\alpha_{min-L}$, $B_{max-L}$, $<B_L>$, $<B_L>S_L$, as well as between $C_{304-F}$ and analogous magnetic field characteristics in follower umbrae. The absolute value of the linear correlation coefficient for all those dependences is no larger than $0.2$. Any correlation is also absent between contrast asymmetry characteristic in the $He~II~304~$\AA$~$ line, $C_{304-L}/C_{304-F}$, and the relations $\alpha_{min-L}/\alpha_{min-F}$, $<\alpha_L>/<\alpha_F>$, $B_{max-L}/B_{max-F}$, $<B_L>/<B_F>$, $<B_L>S_L/<B_L>S_F$. 

At the same time, a positive correlation has been found between $He~II~304~$\AA$~$ contrast above leader umbrae, $C_{304-L}$, and umbral area-averaged angle between magnetic field direction and radial direction at field measurement point, $<\alpha_L>$. The correlation increases if spots with $C_{304-L}>50$ are excluded from the analysis, Fig. 11(A). This plot was based on spots with $<\alpha_L>~<~<\alpha_F>$. At the same time, any correlation between $C_{304-F}$ and $<\alpha_F>$ is absent for followers. Nevertheless, the pattern of the dependence in Fig. 11(A) is indirectly consistent with results in Fig. 10(A). It follows from Fig. 11(A) that contrast $C_{304-L}$ increases, on average, as $<\alpha_L>$ grows. It follows from Fig.10(A) that contrast is higher, on average in follower umbrae compared to leaders. However, it is in followers that the sample-averaged value of $<\alpha_{F-av}>$ is larger than the corresponding value of $<\alpha_{L-av}>$ for leaders.

For cases when $\alpha_{min-L}\le \alpha_{min-F}$ and $(C_{304-L}/C_{304-F})<2$, there is a correlation between $C_{304-L}/C_{304-F}$ and $l_L/l_F$, Fig.8(B), where $l_L$ is the length of the magnetic field line from the leader umbra to its apex, and $l_F$ is the length of the field line from the follower to its apex. In this case, the values of $l_L$ and $l_F$ were averaged for all field lines that it was possible to trace from a leader or a follower. 

It was noted above that there is no relationship between some photospheric magnetic field characteristics in leaders and followers and $He~II~304~$\AA$~$ contrast in these spots. It could be more correct to compare UV intensity contrast to magnetic field properties at heights where $He~II~304~$\AA$~$ emission forms, not in the photosphere, where magnetic field is measured. We plan to perform such work using various methods for magnetic field calculations in the solar atmosphere. It should also be taken into account that $He~II$ ion emission blends with silicon ion emission from the corona at the same wavelength.

\begin{figure*}
 \centering
 \includegraphics[width=0.95\textwidth]{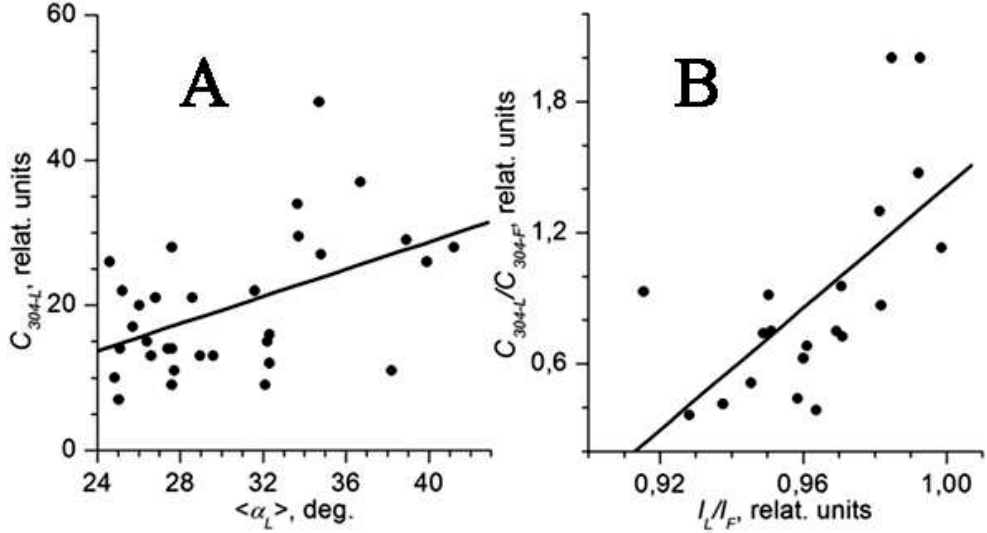}
 \caption{(A) - $\lambda~304~$\AA$~$ contrast $C_{304-L}$ vs angle $<\alpha_L>$ for $C_{304-L}$ below 50 ($C_{304-L}(<\alpha_L>)=0.94<\alpha_L>-8.8$, $r=0.48$); (B) - $C_{304-L}/C_{304-F}$ vs $l_L/l_F$ for spot groups with $C_{304-L}/C_{304-F}\le 2$ and $l_L/l_F\le 1$ ($C_{304-L}/C_{304-F}(I_L/I_F)=13.93I_L/I_F-12.52$, correlation coefficient $r=0.65$).}
\end{figure*}

The physical cause of $He~II~304$\AA$~$ contrast differing between leaders and followers and how this difference is related to magnetic field asymmetry in the umbrae of the two types of spots can be understood by analysing the $He~II~304$\AA$~$ line formation mechanism. According to \citet{Zirin1975}, $He~II~304$\AA$~$ glow in coronal holes is due to the ionisation-recombination mechanism, when helium atoms are ionised by $\lambda\le 228$\AA$~$ emission. {\bf The number of $He~II$ ions on which the radiant flux in the $304$\AA$~$ line depends is determined by ionization equilibrium when velocities of ionization of $He~I$ atoms defined by the shorter-wave coronal emission/radiation flux and the recombination of $He~II$ ions dependent on electron density in the place of the emission generation in the $304$\AA$~$ line are compared. In this case, according to \citet{Zirin1975}, the following relation is valid: $N(He~II)=const~N(He~I)\times (1/n_e)$, where $n_e$ - electron density, $N(He~I)$ - the $He~I$ atom concentration.} Thus, with the $He~I$ atom number being constant (e.g., inside a coronal hole observable in $He~I~10830$\AA) the places of the highest $He~II~304$\AA$~$ intensity will be located in places with the lowest electron density. {\bf We can assume that this mechanism works not only in coronal holes according to \citet{Zirin1975}, but also in sunspots. In followers, magnetic tube field line divergence from the tube axis, which, as a first approximation, can be characterized by $<\alpha>$, is larger than in leaders. This results in lower electron density, at same heights above follower umbrae compared to leader umbrae, and, correspondingly, in brighter $He~II~304$\AA$~$ emission in followers compared to leaders.}
 
This mechanism may also work in sunspots. In followers, magnetic tube field line divergence from the tube axis, which, as a first approximation, can be characterised by $<\alpha>$, is larger than in leaders. This results in lower electron density, at same heights above follower umbrae compared to leader umbrae, and, correspondingly, in brighter $He~II~304$\AA$~$ emission in followers compared to leaders.

\subsection{Discussion and conclusions}

This paper confirms the conclusion in \citep{Zagainova2015, Zagainova2015a} that magnetic properties differ between the umbrae of magneto-conjugated leaders and followers, implying an asymmetry in magnetic properties between leaders/followers. Earlier analyses of magnetic properties in umbrae in \citep{Zagainova2015a} employed high spatial resolution field vector measurements by the SDO/HMI instrument. In this paper, the magnetic properties were also found using the SDO/HMI instrument, but a more accurate and faster azimuth disambiguation of vector magnetograms was used \citep{Rudenko2014}. It has been established that in $\approx 78\%$ of the leading/following sunspot pairs under analysis, the minimum angle between magnetic field direction and the positive (i.e. anti-sunward) normal to solar surface at field measurement point is smaller in leading sunspots compared to following, $\alpha_{min-L}<\alpha_{min-F}$. It has been found that, in $\approx 83\%$ of the spot groups under study, a similar inequality holds for umbral area-averaged angles $<\alpha_{min-L}>~<~<\alpha_{min-F}>$. In other words, for many magneto-conjugated spots, magnetic field in leaders is chiefly closer to the radial direction from the centre of the Sun than in followers.
 
According to SDO data, the average value of $\alpha_{min-L-av}$, obtained by averaging for all leaders under study, was found to be $\sim~2.71$ times smaller than the average value of $\alpha_{min-F-av}$ for followers. According to data in \citep{Zagainova2015} this ratio is $\sim~1.8$. Our analysis of sunspot magnetic properties based on SDO data has also shown that the value of $\alpha_{min}$ in single sunspots with developed umbra and penumbra and pores is smaller than in leadings. For umbral area-averaged angles $<\alpha>$, the ratio ${<\alpha_{L-av}>}/{<\alpha_{F-av}>}$ is approximately $0.85$. At the same time, according to SDO data, the average value of the minimum angle $\alpha_{min-av}$ is smaller in leadings and followings compared to the respective angles found from calculations of these angles using potential approximation field calculations. The conclusion in our previous paper \citep{Zagainova2015, Zagainova2015a} regarding the positive correlation existing between $\alpha_{min-L}$ and $\alpha_{min-F}$ has also been confirmed. In this paper, we show for the first time that a positive correlation also exists between umbral area-averaged values of angle $\alpha$ in leaders and followers: ${<\alpha_L>}$ and $<\alpha_F>$. 

We pointed out in 3.1 that our findings, according to which $\alpha_{min-L}<\alpha_{min-F}$ and $<\alpha_L>~<~<\alpha_F>$ for most of the magneto-conjugated leader/follower pairs we studied, conflicts with theoretical calculations for a magnetic tube ascending from the convection zone base towards the solar surface, as well as with some observations of AR properties. In 3.1, we discussed possible causes of this contradiction. The likeliest way to resolve the contradiction is to suggest that magneto-conjugated leaders and followers should not result from a magnetic tube emerging from the convection zone base. Possible formation mechanisms for such spots are discussed in 3.1.

It was noted in our previous papers that there is practically no correlation between $\alpha_{min}$ and umbral area $S$ in both types of spots (leder/follower). It is shown in this paper that there is a negative correlation between both the minimum angle $\alpha_{min}$ and $S$ in both types types of spots, as well as between umbral area-averaged values of angles $<\alpha>$ and $S$. The largest absolute value of the correlation coefficient ($|r|\approx 0.44$) has been found for the dependence of $<\alpha>_L$ on $S_L$ for leaders. We have confirmed our earlier conclusion in \citep{Zagainova2015a} that a correlation exists between the ratios $\alpha_{min-L}/\alpha_{min-F}$ and $S_L/S_F$, as well as demonstrating the presence of a relationship between $<\alpha_L>/<\alpha_F>$ and $S_L/S_F$. 

A similar conclusion was made in our earlier papers regarding the dependence between maximum magnetic induction $B_{max}$ and angle $\alpha_{min}$ in both types of spots: as $\alpha_{min}$ increased, $B_{max}$ grew, on average, very slowly, while either any correlation between these two values was absent or it was very weak and negative. In this paper, a noniceable negative correlations has been found between $B_{max-L}$ and $\alpha_{min_L}$, as well as between $<B_L>$ and $<\alpha_L>$ (for the latter dependence, $|r|\approx 0.58$) in leaders, with practically no correlation between $B_{max-F}$ and $\alpha_{min_L}$, as well as between $<B_F>$ and $<\alpha_F>$ in followers ($r=-0.115$ and $r=-0.148$). Note that all the above dependences were obtained for spots satisfying this condition: $\alpha_L\le \alpha_F$ or $<\alpha_L>\le <\alpha_F>$. 

In this paper, as in our previous papers \citep{Zagainova2015, Zagainova2015a}, the dependence of the maximum and mean value of magnetic induction on sunspot umbral area are compared, separately for leaders and followers. The major results of our analysis of these dependences can be formulated as follows: 1) neither the maximum no the mean value of magnetic field drop to zero as the area decreases to very small values; 2) in all cases, magnetic field in leading and single spots is larger than in followings; 3) our approximation yields these hold values of the area when the $B_{max}(S)$ and $<B>(S)$ curves begin to be saturated, MSH for leadings and MSH for followings. This corresponds to sunspot umbral radii of 5000 km and 3100 km. The values we obtained agree, on the whole, with estimates by Solov'ev and Kirichek \citep{Solovev2014} and support the concept they develop of a ,,not deep'' sunspot. Thus interpreted, our data indicate that followings may be less deep formations.

It was demonstrated in \citep{Zagainova2011, Zagainova2015a} that, on average, $He~II~304~$\AA$~$ intensity contrast ($C_{304}$) was higher above follower umbrae than above leaders and single spots and only weakly depended on umbral area. \citet{Zagainova2011} suggested that this difference in contrast between the two types of spots is due to different magnetic properties between leader and follower umbrae, or, in other words, due to asymmetry of magnetic properties for such spots. After we first discovered this asymmetry in magnetic characteristics between leaders and followers, the question arose: is there a quantitative relationship between $He~II~304~$\AA$~$ contrast and the magnetic properties of leader and follower umbrae, or between the asymmetry in contrast between the two types of spots and the asymmetry in their magnetic properties. We failed to find any such relationship in \citep{Zagainova2015a}, but in this paper we used a more accurate and faster azimuth disambiguation of vector magnetograms \citep{Rudenko2014} and successfully demonstrated that spots satisfying certain conditions exhibit a positive correlation between $C_{304-L}$ and $<\alpha_L>$ for leaders and between $C_{304-L}/C_{304-F}$ and $l_L/l_F$. Here, $l_L(l_F)$ is the length of a magnetic field line from leader (follower) to the field line apex. For other dependences between $C_{304-L}$, $ C_{304-F}$, $C_{304-L}/C_{304-F}$, on the one hand, and magnetic field characteristics in leader/follower umbrae, on the other, we found either a weak positive correlation or no correlation. Thus we may tentatively suggest that there is a relationship between $He~II~304~$\AA$~$ contrast and magnetic properties of leader and follower umbrae. To obtain dependences with a higher coefficient for correlations between $He~II~304~$\AA$~$ contrast above the umbrae of the two types of spots and magnetic characteristics of the umbrae, one may need to compare $C_{304-L}$, $ C_{304-F}$, $C_{304-L}/C_{304-F}$ to magnetic field properties, not at photospheric level, where the field is measured, but at higher altitudes. 

This paper was partially supported by RFBR grants N 14-02-00308 and N 15-02-01077. The authors are grateful to the SOLIS, SDO/HMI and SDO/AIA teams for making the data of these instruments freely available. The authors thank Rudenko G.V. for the software used to select magneto-conjugated sunspot pairs. The authors are grateful to Rudenko G.V. and Afinogenov S.A. for their software for determining magnetic field components based on SDO/HMI vector magnetograph data, relying on a very fast and accurate azimuth disambiguation of vector magnetograms. 






\begin{thebibliography}{}

\bibitem[Akhtemov et al.(2014)]{Akhtemov2014}
Akhtemov,~Z.S., Andreeva,~O.A., Rudenko,~G.V., Stepanian,~N.N., Fainstein,~V.G. 2014. Temporal variations in the large-scale magnetic field of the solar atmosphere at heights from the photosphere to the source surface. Bull. Crimean Astrophys. Observatory. 110, 108--118. http://dx.doi.org/10.3103/S0190271714010033.

\bibitem[Antalova(1991)]{Antalova1991}
Antalova,~A., 1991. The relation of the sunspot magnetic field and penumbra-umbra radius ratio. Astronomical Institutes of Czechoslovakia, Bulletin (ISSN 0004-6248). 42, 316--320. 

\bibitem[Baranov(1974)]{Baranov1974}
Baranov,~A.V., 1974. Magnetic fields in small sunspots. Astron. circular. 847, 5--6 (in Russian). 

\bibitem[Baumann et al.(2004)]{B04}
Baumann,~I., Schmitt,~D., Sch\"ussler,~M., Solanki,~S.K., Evolution of the large-scale magnetic field on the solar surface: a parameter study, Astron. Astrophys., 426, 1075--1091, 2004.

\bibitem[Beckers and Schroter(1968)]{Beckers1968}
Beckers,~J.M., Schroter,~E.H., 1968. The intensity, velocity and magnetic structure of a sunspot region. I: Observational technique; properties of magnetic knots. Solar Phys. 4, 142--164. http://dx.doi.org/10.1007/BF00149561.

\bibitem[Bray and Loughhead(1964)]{Bray1964}
Bray,~R., Loughhead,~R., 1964. Sunspots. Volume Seven. The International Astrophysics Series. Chapman and Hall LTD. London.

\bibitem[Bumba(1967)]{Bumba1967}
Bumba,~V., 1967. Magnetic fields in small and young sunspots. Solar Phys. 1, 371--376. http://dx.doi.org/10.1007/BF00151362

\bibitem[Caligari et al.(1995)]{Caligari1995}
Caligari,~P., Moreno-Insertis,~F., Schussler,~M., 1995. Emerging flux tubes in the solar convection zone. I. Asymmetry, tilt, and emergence latitude. Astrophys. J. 441, 886--902. http://dx.doi.org/10.1086/175410.

\bibitem[Cowling,(1946)]{Cowling1946}
Cowling,~T.G., 1946. The growth and decay of the sunspot magnetic field, Mon. Not.Roy. Astron. Soc. 106, p.218--224. http://adsabs.harvard.edu/abs/1946MNRAS.106..218C 

\bibitem[Fan et al.(1993)]{Fan1993}
Fan,~Y., Fisher,~G.H., DeLuca,~E.E., 1993. The origin of morphological asymmetries in bipolar active regions. Astrophys. J. 405, 390--401. http://dx.doi.org/10.1086/172370.

\bibitem[Gilman and Howard(1985)]{Gilman1985}
Gilman,~P.A., Howard,~R., 1985. Rotation rates of leading and following sunspots. Astrophys. J. 295, 233--240. http://dx.doi.org/10.1086/163368.

\bibitem[Harvey and Livingston(1969)]{Harvey1969}
Harvey,~J., Livingston,~W., 1969. Magnetograph measurements with temperature-sensitive lines. Solar Phys. 10, 283--293. http://dx.doi.org/10.1007/BF00145515.

\bibitem[Harvey(1971)]{Harvey1971}
Harvey,~J., 1971. Solar magnetic fields - small scale. Publ. Astron. Soc. Pac. 83, 539--549. http://dx.doi.org/10.1086/129171.

\bibitem[Houtgast and van Sluiters(1948)]{Houtgast1948}
Houtgast,~J., Van.Sluiters,~A., 1948. Statistical investigations concerning the magnetic fields of sunspots, I. Bull. Aston. Inst. Netherlands. 10, 325--333.

\bibitem[Jin et al.(2006)]{Jin2006}
Jin,~C.L., Qu,~Z.Q., Xu,~C.L., Jhang,~X.Y., Sun, M.G., 2006. The relationships of sunspot magnetic field strength with sunspot area, umbral area and penumbra-umbra radius ratio. Astrophys. Space Sci. 306, 23--27. http://dx.doi.org/10.1007/s10509-006-9217-6.

\bibitem[Kuklin(1985)]{Kuklin1985}
Kuklin,~G.V., 1985. East-west asymmetry in the Wilson effect. Researches in geomagnetism, aeronomy and physics of the Sun. 73, 52--60 (In Russian).

\bibitem[Lemen et al.(2012)]{Lemen2012}
Lemen,~J.R., Title,~A.M., Akin,~D.J., et al., 2012. The atmospheric imaging assembly (AIA) on the Solar Dynamics Observatory (SDO). Solar Phys. 275, 17--40. http://dx.doi.org/10.1007/s11207-011-9776-8.

\bibitem[Livingston and Harvey(1969)]{Livingston1969}
Livingston,~W., Harvey,~J., 1969. Observational evidence for quantization in photospheric magnetic flux. Solar Phys. 10, 294--296. http://dx.doi.org/10.1007/BF00145516.

\bibitem[Livshits(1975)]{Livshits1975}
Livshits,~M.A., 1975. Constancy of tau /10830/ in plages and helium emission in a shortwave-radiation field. Astron. Zh. 52, 970--974 (in Russian).

\bibitem[Maltby(1992)]{Maltby1992}
Maltby,~P., 1992. Continuum observations and empirical models of the thermal structure of sunspots. Proc. NATO Advanced Research Workshop on the Theory of Sunspots, Cambridge, United Kingdom. Sept. 22-27. (A93-47383 19-92). 103--120. 

\bibitem[Murdin(2000)]{Murdin2000}
Murdin,~P., 2000. Sunspot Magnetic Fields Encyclopedia of Astronomy and Astrophysics. Edited by Paul Murdin, article 2298. Bristol: Institute of Physics Publishing. 

\bibitem[Nikolskaya(1966)]{Nikolskaya1966}
Nikolskaya,~K.I. 1966. He I Excitation in chromospheric spicules Astron. Zh. 43, 936 (in Russian).

\bibitem[Obridko(1985)]{Obridko1985}
Obridko,~V.N., 1985. Sunspots and complexes of activity. M.: Nauka, 256 pp. (in Russian).

\bibitem[Obridko and Shelting(2013)]{Obridko2013}
Obridko,~V.N., Shelting,~B.D., 2013. Global complexes of activity. Astronomy Reports, 57, 786--796. http://dx.doi.org/10.1134/S1063772913100041.

\bibitem[Pipin and Kosovichev(2011)]{Pipin2011}
Pipin,~V.V., Kosovichev,~A.G., 2011. The subsurface-shear-shaped solar $\alpha$ dynamo Astrophys. J. 727, 1--4. http://dx.doi.org/10.1088/2041-8205/727/2/L45.

\bibitem[Pozhalova(1988)]{Pozhalova1988}
Pozhalova,~Zh. A., 1988. The study of selected helium lines in the solar spectrum. Astron. Zh. 65, 1037--1046 (in Russian).

\bibitem[Ringnes and Jensen(1960)]{Ringnes1960}
Ringnes,~T.B., Jensen,~E., 1960. On the relation between magnetic fields of sunspots in the interval 1917-56. Astrophysica Norvegica. 7, 99--121. 

\bibitem[Rudenko(2001)]{Rudenko2001}
Rudenko,~G.V., 2001. Extrapolation of the solar magnetic field within the potential-field approximation from full-disk magnetograms. Solar Phys. 198. 5--30. http://dx.doi.org/10.1023/A:1005270431628.

\bibitem[Rudenko and Anfinogentov(2014)]{Rudenko2014}
Rudenko,~G.V., Anfinogentov,~S. A., 2014. Very Fast and Accurate Azimuth Disambiguation of Vector Magnetograms. Solar Phys. 289. 1499-1516. http://dx.doi.org/10.1007/s11207-013-0437-y.

\bibitem[Scherrer et al.(2012)]{Scherrer2012}
Scherrer,~P.H., Schou,~J., Bush,~R.I., Kosovichev,~A.G., Bogart,~R.S., Hoeksema,~J.T., Liu,~Y., Duvall,~T.L., Zhao,~J., Title,~A.M., Schrijver,~C.J., Tarbell,~T.D., Tomczyk,~S. Solar Phys., V. 275, Issue 1-2, 207--227. http://dx.doi.org/10.1007/s11207-011-9834-2.

\bibitem[Sheeley(1966)]{Sheeley1966}
Sheeley,~N.R.,~Jr., 1966. Measurements of solar magnetic fields. Astrophys. J. 144, 723--732. http://dx.doi.org/10.1086/148651.

\bibitem[Sheeley(1967)]{Sheeley1967}
Sheeley,~N.R.,~Jr., 1967. Observations of small-scale solar magnetic fields. Solar Phys. 1, 171--179. http://dx.doi.org/10.1007/BF00150852.

\bibitem[Sobotka(1986)]{Sobotka1986}
Sobotka,~M., 1986. Semi-empirical models of sunspots in various phases of evolution Contributions of the Astronomical Observatory Skalnate Pleso. 15, 315--318. 

\bibitem[Solovev and Kirichek(2014)]{Solovev2014}
Solovev,~A., Kirichek,~E., 2014. Basic properties of sunspots: equilibrium, stability and long-term eigen oscillations Astrophys. Space Sci. 352, 23--42. http://dx.doi.org/10.1007/s10509-014-1881-3.

\bibitem[Stenflo(1973)]{Stenflo1973}
Stenflo,~J.O., 1973. Magnetic-field structure of the photospheric network. Solar Phys. 32, 41--63. http://dx.doi.org/10.1007/BF00152728.

\bibitem[Steshenko(1967)]{Steshenko1967}
Steshenko,~N.V., 1967. Magnetic fields in small sunspots and pores. Izv. Krym. astrophys. obs. 37, 2126 (in Russian). 

\bibitem[Van Driel-Gesztelyi and Petrovay(1990)]{Van Driel-Gesztelyi1990}
Van Driel-Gesztelyi,~L., Petrovay,~K., 1990. Asymmetric flux loops in active regions, I. Solar Phys. 126, 285--298. http://dx.doi.org/10.1007/BF00153051.

\bibitem[Zagainova(2011)]{Zagainova2011}
Zagainova,~Yu.S., 2011. He II 304 emission above sunspot umbrae. Astronomy Reports. 55, 159--162. http://dx.doi.org/10.1134/S1063772910121030.

\bibitem[Zagainova et al.(2015)]{Zagainova2015}
Zagainova,~Yu.S., Fainshtein,~V.G., Rudenko,~G.V., Obridko,~V.N., 2015. A comparative analysis of magnetic field properties in leading and following sunspots. Astronomy Reports, 59, 156--164. http://dx.doi.org/10.1134/S1063772914120129.

\bibitem[Zagainova et al.(2015a)]{Zagainova2015a}
Zagainova,~Yu.S., Fainshtein,~V.G., Obridko,~V.N., 2015a. Comparison of the properties of leading and trailing sunspots. Geomagnetism and Aeronomy, 55. 13–23 (in Russian). http://dx.doi.org/10.1134/S001679321406022X.

\bibitem[Zirin(1975)]{Zirin1975}
Zirin,~H. 1975. The helium chromosphere, coronal holes, and stellar X-rays. Astrophys. J., 199. L63--L66. http://dx.doi.org/10.1086/181849.

\end{thebibliography}


\end{document}